\documentclass{vgtc}                          

\graphicspath{{figures/}{pictures/}{images/}{./}} 

\usepackage{times}                     

\usepackage{tabu}                      
\usepackage{booktabs}                  
\usepackage{lipsum}                    
\usepackage{mwe}                       

\usepackage{mathptmx}                  

\onlineid{0}

\vgtccategory{Research}

\vgtcinsertpkg

\title{LibraryLens: An Interactive Tool for Exploring and Arranging Digital Bookshelves}

\author{Trevor DePodesta\thanks{e-mail: tdepodesta@seas.harvard.edu} %
\and Johanna Beyer\thanks{Dr. Johanna Beyer, of blessed memory, contributed to the conceptual framework and early development of this research, but passed away before manuscript completion. We include her as an author to honor her intellectual contributions, while noting she did not review the final submission.}}
\affiliation{\scriptsize Harvard University}

\teaser{
  \centering
  \includegraphics[width=.85\linewidth]{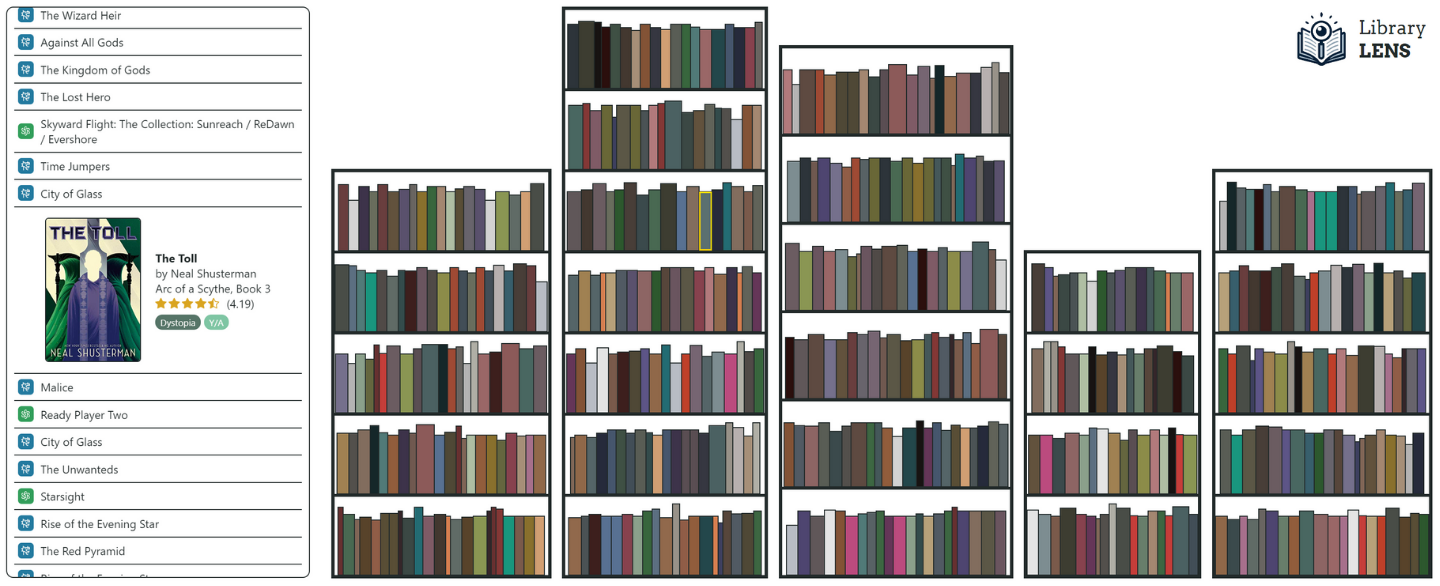}
  \caption{LibraryLens rendering a collection of 650+ books on a set of user defined custom bookshelves.}
  \label{fig:teaser}
}

\abstract{
    Existing digital book management platforms often fail to capture the rich spatial and visual cues inherent to physical bookshelves, hindering users' ability to fully engage with their collections. We present LibraryLens, a novel visualization tool that addresses these shortcomings by enabling users to create, explore, and interact with immersive, two-dimensional representations of their personal libraries. The tool also caters to the growing trend of social sharing within online book communities, allowing users to create visually appealing representations of their libraries that can be easily shared on social platforms. Despite limitations inherent to the metadata being rendered, formative evaluations suggest that LibraryLens has the potential to lower the barrier to entry for users seeking to optimize their book organization without the constraints of physical space or manual labor, ultimately fostering deeper engagement with their personal libraries.
} 


\begin{document}


\firstsection{Introduction}

\maketitle

In the evolving landscape of digital media, the charm of a physical book collection still persists, offering a tangible connection to the literature that digital alternatives struggle to replicate. Personal book collections represent a uniquely tactile and visually rich domain where digital tools have largely failed to provide a satisfying analog to the experience of managing and interacting with physical books. As these collections grow however, the tasks of organizing, visualizing, and interacting with them becomes increasingly taxing and complex. Current digital management solutions such as Goodreads and BookBuddy offer robust tools for cataloging and tracking reading habits, but fall short in providing an immersive and intuitive representation of users' libraries \cite{Goodreads, BookBuddy}. These platforms default to one-dimensional lists or grids to convey library metadata, violating users' mental models of how such data is traditionally physically represented and significantly reducing the rich spatial and visual context that physical bookshelves afford. This disconnect between digital and physical representations can hinder users' ability to effectively explore, organize, and engage with their collections.

Even in physical environments, one of the primary challenges faced by book collectors and enthusiasts is the limited space available for arranging and displaying their collections. As personal libraries grow, it becomes increasingly difficult to physically experiment with different organization schemes and layouts. Rearranging dozens, hundreds, or even thousands of books to test the aesthetic feasibility of various permutations is a time-consuming and labor intensive task. Moreover, many individuals may not have the luxury of dedicated space to spread out their books and asses the visual impact of each arrangement. This constrains book lovers' ability to explore and optimize their library organization, ultimately hindering their engagement with and enjoyment of their collections.

Addressing these gaps, we introduce LibraryLens, a visualization tool that aims to bridge the gap between digital and physical book organization, allowing users to create a more natural and immersive representation of their personal libraries by generating an interactive, two-dimensional representation of their physical bookshelves.

\section{Related Work}
\subsection{Existing Book Cataloging Tools}
While commercially tailored tools for library cataloging efforts have been around for half a century \cite{ILS, OCLC}, over the past decade and a half several digital platforms have emerged to help lay book enthusiasts catalog and manage their personal libraries. Goodreads, arguably the most popular social book cataloging website with over 100 million users, allows users to register books to generate library catalogs and reading lists \cite{Goodreads}. However, the platform primarily focuses on book discovery and recommendation, with limited options for visualizing and organizing personal collections beyond a simple "shelf" metaphor.

Similarly, LibraryThing is a web-based platform for storing and sharing personal library catalogs \cite{LibraryThing}. While it offers classification features, its visualization options are limited to a basic gallery view of book covers. Notably, LibraryThing---along with most other popular personal library management tools---encourages users to upload their collections by scanning ISBN barcodes \cite{LibraryThing, libib, BookBuddy, Goodreads}. Overall, these platforms primarily focus on cataloging and management, with less emphasis on the visual and spatial aspects of book collections that are central to the physical library experience.

\subsection{Reading as a Social Experience}
The largest (non-commercial) book cataloging platforms double as social media platforms for users to both share and view their respective collections \cite{Goodreads, LibraryThing}, with a 2017 study revealing that Goodreads has become a hybrid social navigation site where users do not often ignore either the social or book activities on the platform at the expense of the other \cite{Thelwall2017Goodreads}. On these sites, users benefits not just from the official, boilerplate book metadata provided to them, but also from the social metadata: comments, reviews, tags, ratings, etc. Social Book Search (SBS) research aims to exploit this social metadata to group related texts (often in the context of recommendations), but the ambiguity of natural language queries remains an issue \cite{Ullah2020Social}.

The rise of social media has led to the emergence of "bookish" communities on platforms such as YouTube (BookTube), TikTok (BookTok), and Instagram (Bookstagram) \cite{reddan2024social, bookTube/Tok, selfies_shelfies}. In these spaces, readers share their love for books, discuss their reading experiences, and showcase their personal libraries. A recent study on engagement with BookTube and BookTok influencers revealed that the three primary motivations for reading among these communities are achievement, escapism, and social interaction \cite{bookTube/Tok}. Readers view their shelves not only as storage repositories but as trophy cases that display their accomplishments, using hashtags like \#bookstagram and \#shelfie to share these experiences with others \cite{selfies_shelfies}. However, in a trend that feeds into overconsumerism, a lack of a visually robust book collection could alienate readers who feel pressure to keep up with the aesthetics and quantity of books showcased by others in these communities.

\subsection{Aligning with Mental Models}
The concept of mental models with respect to information visualization refers to the internal representations of external visualization systems. Liu and Stasko find that mental models preserve not just item level information about the data at hand, but also schematic information \cite{mentalModels_infoVis}. This means that users of the platforms above (who reasonably have prior experience with physical bookshelves) have established mental models about how book collection data is normally conveyed: two-dimensional shelves. Creating visualizations that align with users' existing mental models helps to facilitate deeper cognitive engagement with the data at hand, which is why LibraryLens preserves the 2D shelf view in a digital medium \cite{mentalModels_infoVis, Rapp2005}.

\subsection{Browsable Visualization as a Discovery Tool}
While the existing platforms discussed above help facilitate the discovery of \textit{new} titles, LibraryLens aims to encourage discovery of connections between books \textit{within} a collection. Early attempts at visualizing metadata of documents in digital libraries such as libViewer \cite{libViewer} laid the groundwork for more advanced systems like Blended Shelf \cite{blendedShelf}, a 3D interactive visualization for library-goers to browse library collections that weren't on public display. Other tools such as the Bohemian Bookshelf prioritize serendipity, the faculty of making fortunate discoveries accidentally while searching for something else \cite{bohemian}. While some of these tools help to preserve users' spatial mental models \cite{libViewer, blendedShelf}, they do not cater to consumer collections at the scale that LibraryLens targets.

\begin{figure}[tb]
 \centering
 \includegraphics[width=\columnwidth]{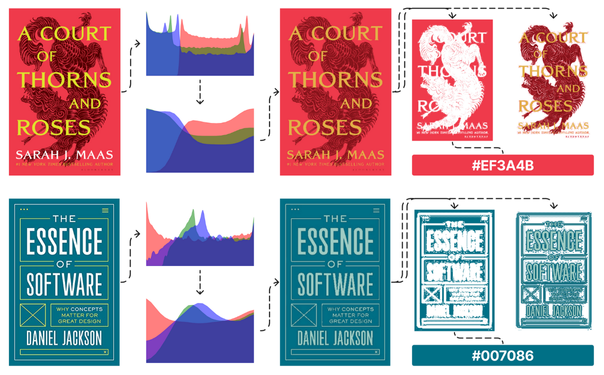}
 \caption{Cover images are broken into pixel arrays, converted to RGB color space, and quantized with restricted palettes (n=4) before the primary remaining color is isolated and returned to render the book spines. The top and bottom RGB histograms reflect (bin=256) and (bin=24), respectively.}
 \label{fig:quantization_pipeline}
\end{figure}

\begin{figure}[tb]
 \centering
 \includegraphics[width=\columnwidth]{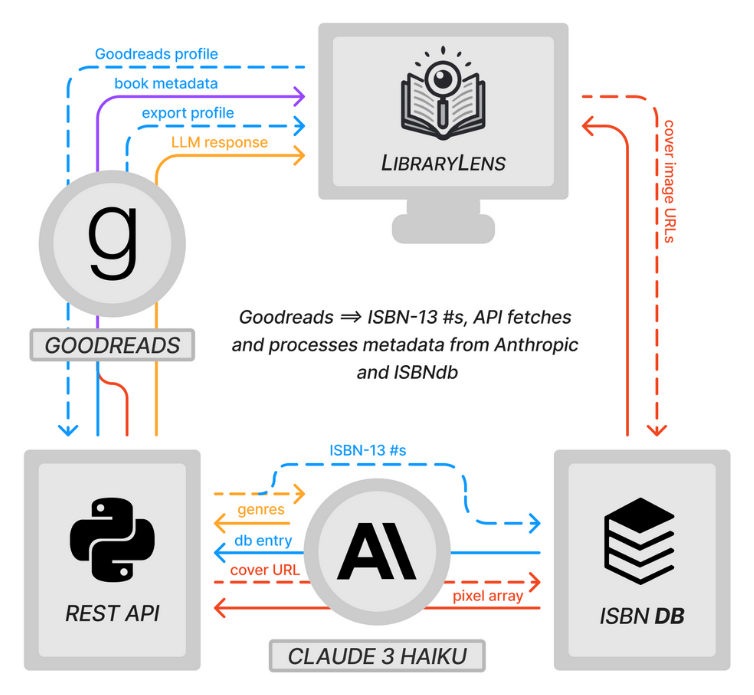}
 \caption{LibraryLens pipeline. User exports ISBNs from Goodreads $\rightarrow$ Raw metadata fetched from ISBNdb $\rightarrow$ Claude normalizes genre + age facets $\rightarrow$ Cover images are loaded and quantized $\rightarrow$ Renderer streams an organized shelf scene to the front-end}
 \label{fig:sys_arch}
\end{figure}

\begin{figure*}[tb]
 \centering 
 \includegraphics[width=\textwidth]{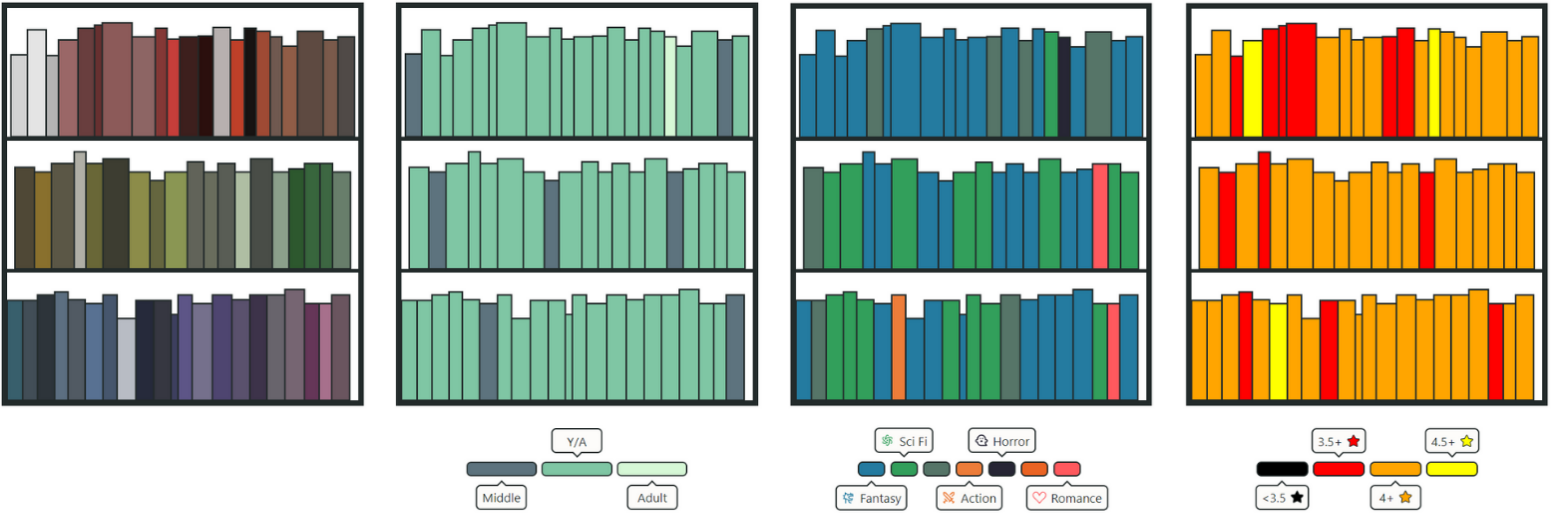}
 \caption{The same shelf with four alternate spine color encodings to understand global patterns at a glance.  From left to right: (a) Original spine art; (b) Reading-age palette exposes a lack of adult literature; (c) Genre coloring reveals overwhelming Fantasy and Science Fiction dominance; (d) Goodreads reviews add a visual dimension to SBS, immediately drawing attention to the highest and lowest rated novels.}
 \label{fig:spine_encoding}
\end{figure*}

\section{LibraryLens}
LibraryLens is a visualization tool designed to create an immersive and interactive representation of personal book collections. The tool allows users to directly input their book data in the form of the exported library CSV file provided to them in the \textit{My Books $\rightarrow$ Import/Export} page of their Goodreads profile \cite{Goodreads}. This file is parsed to reveal an array of ISBN-13s, which are used to retrieve detailed metadata from the ISBNdb database \cite{ISBNdb}. A more detailed information generation workflow can be seen in \cref{fig:sys_arch}.

A key feature of LibraryLens is the algorithmic arrangement of books on virtual shelves based on user-specified dimensions and preferences. The tool takes into account factors such as book size (generally tracked ISBN metadata; exceptionally estimated from page counts), shelf space, and user-defined sorting strategies to create a visually appealing and organized representation of the user's collection. LibraryLens employs a cover image quantization technique (\cref{fig:quantization_pipeline}) to determine the dominant color of each book's spine, enhancing the visual fidelity of the virtual bookshelves.

Users can interact with the visualization by applying various sorting strategies from a control panel on the right of the screen. Collections can be organized by a combination of size, color, alphabetization, author/series, average rating, genre, age, and more. Hover events reveal a more complete picture of the volume's metadata in a side panel on the left, and an additional bottom menu controls the spinal color encoding as illustrated in \cref{fig:spine_encoding}. Additionally, LibraryLens supports direct manipulation of individual books within the visualization through a drag-and-drop interaction---after which the layout re-flows automatically---enabling fine-grained control over default sorted results.

\subsection{Metadata Normalization}
\label{sec:normalization}
Public book databases often aggregate dozens of near-duplicate genre distinctions, making facet-based color palettes visually noisy. Furthermore, these and other tags are occasionally internally inconsistent within series, or otherwise missing entirely. LibraryLens therefore interfaces with a custom built tool in Anthropic's Claude 3 API \cite{claude} that maps (or in the case of missing data, generates) the ISBN record to a small, enumerated set of genres and reading-age bands. Constraining the output space guarantees consistent bucketing and avoids the "million-sub-genre" problem raised by early testers. 

\subsection{Implementation}
LibraryLens is a web application developed in the React framework \cite{react} and the Flask microframework \cite{flask}.

\section{Evaluation}
Our evaluation is based on three illustrative usage scenarios and formative impressions from two informal playtesting sessions with mock data provided to self proclaimed casual and 'super' prospective users.
\subsection{Usage Scenarios}

\subsubsection{Indecisive Organizer}
A user named Alec observes the bookshelves in his bedroom, daunted by the task of manually reorganizing a nearly 400 volume collection. He starts by taking an alphabetical approach, but two shelves in changes his mind, not liking the way the books were looking. With a sigh he starts taking the books down to start over, but quickly runs out of available floor space. Instead, he turns to LibraryLens. An avid reader, Alec is already accustomed to cataloging his books digitally, and simply imports his library to the tool. Here, he can test various sorting permutations within minutes. Changing the spinal color encodings (\cref{fig:spine_encoding}) reveals to him that he doesn't have enough genre diversity to justify a subdivision by genre, so Alec instead opts to sort his books by color. A user with minor color vision deficiency, Alec feels that a few of the books seem out of place, so he manually drags and drops the outliers to their new locations and the visualization automatically adjusts spacing and layout to accommodate. Once he's satisfied with the results, he downloads a labeled rendering to use as a blueprint and returns to his shelves, armed with a plan.

\subsubsection{Informed Browser}
A user named Jayna, seeking a new quick book to read, browses the book lists recommended to her by Goodreads, adding potential titles to her 'Want to Read' shelf on on the platform. Unfortunately, when she goes to refine that list, she discovers it is impossible to use genre or age demographics as filtering tools; the best way to peruse the list is a single book-column table view of a few dozen entries at a time. Frustrated, Jayna exports the shelf and loads it into LibraryLens. There, she can more easily find her next read by sorting candidates by genre and average Goodreads rating, and mentally filters out any titles that would take too long to read with a quick visual scan of the rendered shelves. By leveraging SBS techniques to curate a starting pool and then using LibraryLens to visually refine her choices, Jayna successfully makes a selection.

\subsubsection{Digital Bibliophile}
\label{sec:digital_bibliophile}
A user named Mark is an avid reader of e-books, listener of audiobooks, and active patron of his local library, leading to him possessing a relatively small personal physical book collection. Mark feels left out when he sees trends on \#BookTok and \#bookstagram with his friends sharing their reading accomplishments by posting "shelfies" of meticulously arranged bookshelves. Leveraging the partnership between Amazon and Goodreads, Mark automatically imports his Kindle \cite{kindle} and Audible \cite{audible} books into a Goodreads shelf, and manually searches to add his latest library reads. Feeding that export into LibraryLens, Mark plays around with the bookshelf dimensions and sortings---focusing purely on aesthetics---to generate a visualization that highlights his reading accomplishments in a far more tangible manner that he can then go and share with his friends on social media.

\subsection{Domain User Feedback}

\subsubsection{Informal Playtests}
The first session focused on formative needfinding with three domain experts previewing an early prototype that simply rendered spines for a supplied ISBN list. The second session involved a summative evaluation where a group of students were provided with a sample Goodreads profile containing 135 saved books across seven genres, a subsection of which is visualized in \cref{fig:spine_encoding}. Most tried at least two encodings; several noticed that genre coloring revealed clusters of Middle-Grade Fantasy crammed between Y/A titles, prompting drag-and-drop corrections.

\subsubsection{Key Takeaways}
\textit{Granular control is essential}: Testers accepted automatic layouts but expected manual overrides for edge cases (e.g. series, oversized books).
\textit{Facet-based recoloring surfaces blind spots}: Two testers remarked they hadn't realized how few five-star reads the profile included until the palette changed.
\textit{Consistency beats specificity}: Sortable facets were found to be clean and scannable, thanks to LLM normalization (\cref{sec:normalization}).

The summative session yielded overwhelmingly positive feedback from a broader range of users, revealing an unanticipated group of secondary stakeholders, such as the digital bibliophile (\cref{sec:digital_bibliophile}). Comments on usability and user experience prompted an overhaul of the application's linguistic layer, adding labels and clarifying annotations. Multiple testers reported enjoying the freedom to fine-tune offered by the manual manipulation of sorted results, though some users expressed a desire for the inclusion of mezzo-level interactions (\cref{sec:future_work}).

\section{Discussion}
\begin{table}[tb]
    \scriptsize
    \centering
    \begin{tabu}{%
	r%
	*{2}{c}%
	r%
	*{2}{c}%
	r%
    }
    \toprule
    Genre &   \rotatebox{0}{\# HC} &   \rotatebox{0}{\# SC} &   \rotatebox{0}{Total \#} &   \rotatebox{0}{HC \%} &   \rotatebox{0}{SC \%} &   \rotatebox{0}{\textbf{Total \%}}   \\
    \midrule
    Fantasy & 96 & 42 & 138 & 100\% & 100\% & \textbf{100\%} \\
    Sci Fi & 15 & 17 & 32 & 87\% & 100\% & \textbf{94\%} \\
    Dystopian & 13 & 8 & 21 & 100\% & 100\% & \textbf{100\%} \\
    Mystery & 28 & 8 & 36 & 96\% & 100\% & \textbf{97\%} \\
    Horror & 2 & 11 & 13 & 100\% & 91\% & \textbf{92\%} \\
    Historical & 2 & 8 & 10 & 100\% & 100\% & \textbf{100\%} \\
    Romance & 6 & 4 & 10 & 100\% & 75\% & \textbf{90\%} \\
    Classics & 19 & 48 & 67 & 100\% & 60\% & \textbf{72\%} \\
    Other & 26 & 5 & 31 & 96\% & 80\% & \textbf{94\%} \\
    \midrule
    \textbf{Fiction} & \textbf{207} & \textbf{151} & \textbf{358} & \textbf{98\%} & \textbf{85\%} & \textbf{93\%} \\
    \midrule
    Nonfiction & 113 & 91 & 204 & 86\% & 77\% & \textbf{82\%} \\
    \midrule
    \textbf{Total} & \textbf{320} & \textbf{242} & \textbf{562} & \textbf{94\%} & \textbf{82\%} & \textbf{89\%} \\
    \bottomrule
    \end{tabu}%
    \vspace{1em}
    \caption{Percentage of books with consistent spine and cover colors by genre and print edition: hardcover (HC) and softcover (SC).}
    \label{tab:consistency}
\end{table}

LibraryLens serves as a proof-of-concept design probe, rather than a fully-featured production system. We deliberately implemented a minimal feature slice that supports end-to-end exploration with real data, knowing that the design space is far richer than what a short-paper prototype can cover. We therefore treat the following subsections not simply as “bugs to fix,” but as signposts toward numerous extensions that this early probe has surfaced.

\subsection{Limitations}
While LibraryLens offers a novel approach to visualizing and organizing personal book collections, it is not without limitations. One such limitation is the potential discrepancy between the algorithmically determined spine color and the actual spine color of the physical book. While the method used (\cref{fig:quantization_pipeline}) is described by previous works as "adequate in many cases" \cite{blendedShelf}, we conducted a study by randomly sampling 562 books across various genres and print editions in order to investigate the frequency of this issue.  The results, shown in \cref{tab:consistency}, reveal that the problem is virtually nonexistent in hardcover fiction publications but more prone to occur in softcover nonfiction publications. While more data points are needed to confirm this random sample, it appears that---following our collected distribution---LibraryLens will accurately render spine colors for approximately 9/10 books. 

Another limitation is LibraryLens' reliance on the availability and accuracy of book metadata in existing databases.  Inconsistencies in older titles with various reprints and incompleteness in ultra-contemporary publications can hinder the tool's effectiveness. Currently, LibraryLens does not offer a solution for manual data entry to compensate for missing pieces of data in the ISBNdb database. Scalability is also a concern, as the frontloaded computation and API calls to Claude become too much of a strain on the browser once a user's collection surpasses a few thousand volumes.  Future work will explore more efficient batching mechanisms to reduce client-side wait time.

\subsection{Future Work}
\label{sec:future_work}
To address these limitations an expand the capabilities of LibraryLens, several avenues for future work have been identified. One promising direction is the exploration of advances in computer vision as an input mechanism. Early experiments with GPT 4 Vision \cite{gpt4_vision} show potential for extracting ISBNs by reverse searching the ISBNdb database using only the metadata gleaned by an LLM from uploaded pictures of book spines. However, the accuracy is still far too inconsistent for use in production.

Users have also expressed interest in mezzo-level interactions, where macro interventions are the global sorting operations and micro-level interactions include drag-and-drop functionality. This could involve preserving groups of books like series or sorting within subsections of the dataset. Furthermore, we hope to explore integrating recent developments in steerable clustering of text embeddings \cite{steerable} to allow for natural language-powered organization tasks, enhancing the user experience and providing more intuitive ways to interact with personal book collections.

\section{Conclusion}
LibraryLens has the potential to simplify the process of visualizing and reorganizing personal book collections, providing a veritable digital sandbox for users to experiment with different book organization strategies without the limitations of physical space, a significant sunk time cost, or the need for manual labor. By providing a visually rich and interactive representation of a user's library, our system aims to encourage exploration and rediscovery, and lower the barrier to entry for readers to share their collections.

\acknowledgments{
The authors wish to thank Elena Glassman, Simon Worchol, Martin Wattenberg, and Fernanda Viégas for their valuable input and discussions in conceiving and refining this work. Special thanks to Alec Sheres, Scott Rifkin, and Sean Kelly for helping to inspire this work.}

\bibliographystyle{abbrv-doi}

\bibliography{template}
\end{document}